# The balance of evidence indicates a slowing AMOC

Stefan Rahmstorf[1,2], Levke Caesar[1]

[1]Potsdam Institute for Climate Impact Research, Potsdam, 14412, Germany
[2]University of Potsdam, Potsdam, 14469, Germany

**Abstract.** The Atlantic Meridional Overturning Circulation (AMOC) is a critical component of the climate system. Climate models have long predicted a slowing of the AMOC due to anthropogenic forcing, with a risk of passing tipping points in the future. The question of whether the AMOC is already weakening is still somewhat controversial. Continuous monitoring is only available for the past two decades, a period too short to tease apart contributions of natural variability and climate change. Reconstructions of AMOC strength going back further in time have different limitations which are debated in the literature. Here we compile and discuss this heterogeneous body of evidence, focusing on latitudes south of 45°N where northward and southward flow are separated vertically. We conclude that the balance of multiple lines of evidence strongly supports a past and ongoing AMOC slowing in response to global warming, although some uncertainty remains regarding its magnitude and timing.

## 1 Introduction

The AMOC[1] is a crucial and sensitive part of the global climate system, yet its recent evolution remains the subject of persistent debate. While some studies report a substantial long-term weakening (e.g. (Caesar et al., 2018; Thornalley et al., 2018)), others have argued that no significant long-term trend can be identified (e.g. (Worthington et al., 2021; Terhaar et al., 2025)). This apparent contradiction has led to a widespread perception that the observational evidence for AMOC change is inconclusive. Here, we argue that this perception is wrong. The available evidence is far more consistent than is generally appreciated; many of the apparent differences relate to different time intervals or regions, or the interpretation of trends not quite reaching the conventional 95% significance level.

Here, we focus on longer time scale trends (> 20 years) rather than short-term variability. We also limit our analysis to the AMOC south of 45°N, where northward heat transport peaks and north- and southward flow are clearly separated vertically, whilst further north they are separated horizontally (Foukal and Chafik, 2024). Because of stable density stratification and little water-mass transformation in the subtropical and tropical ocean, for flow continuity reasons one may expect a coherent overturning not unlike Broecker's metaphor of a 'conveyor belt' here, which however breaks down north of 45°N. There the circulation becomes more complex: deep convection and water-mass transformation occur, the circulation branches towards different formation regions, and some deep water recirculates northward. High-latitude northern components can even anticorrelate with the AMOC farther south (Roewer et al., 2026). In the following sections we will examine different lines of evidence based on a range of ocean variables.

[1] The terms thermohaline circulation (THC) and Atlantic Meridional Overturning Circulation (AMOC) have historically often been used interchangeably, but they refer to different concepts. THC describes the component of the global overturning circulation driven by density differences arising from temperature and salinity, whereas AMOC refers specifically to the Atlantic overturning circulation as diagnosed from the meridional stream function, which includes both buoyancy-driven and wind-driven components. In this paper, we use AMOC in this broader dynamical sense.

Climate models robustly project a weakening of the AMOC in response to anthropogenic warming (Intergovernmental Panel on Climate Change, 2021), raising the question of whether such a decline is already detectable in observations. Continuous measurements from the RAPID AMOC array at 26°N exist only since 2004 (McCarthy et al., 2025), and indeed they show a weakening trend (Figure 3) which is consistent with climate model projections (Drijfhout et al., 2025). Our analysis of the corrected RAPID data (Volkov et al., 2024), for April 2004 to March 2024 gives a trend of −0.95 Sv per decade, or about **1.9 Sv** overall, consistent with McCarthy et al. (2025). While this rate may appear small, at this rate the AMOC could decline from its average ~17 Sv during the past two decades down to ~10 Sv by 2100, as climate models indeed suggest (Portmann et al., 2026) and which is close to the threshold where AMOC is likely already past its tipping point and on the way to shutdown (Drijfhout et al., 2025).

Notably, observations from four mooring arrays from 16.5°N to 42.5°N, focusing on the deep western overturning transport, show a meridionally consistent and highly significant weakening trend since the early 2000s. At 26°N, the decline is 2.6 ± 0.7 Sv per decade over 2004-2023, corresponding to an overall weakening from 20 Sv down to 15 Sv (Xing et al., 2026). These observations do not measure the complete basin-wide AMOC, but they document a weakening of a physically important component of the overturning circulation.

But even more than twenty years of measurements at the RAPID mooring array are still too short to be sure to what extent the observed slowing is natural variability and to what extent a response to global warming. That is why many studies have been looking at possible 'AMOC fingerprints' in ocean properties for which we have much longer measurements, often based on temperature and salinity but also other relevant data. Here we discuss what these studies have found, to consider the balance of evidence provided by a range of different approaches with different limitations and uncertainties.

## 2. The appearance of the 'cold blob'

The concern about whether the AMOC is already slowing started in earnest with the appearance of a 'warming hole' or 'cold blob' in the northern Atlantic. One of us mentioned this already in a 1995 newspaper article as a possible sign of a slowing AMOC (Rahmstorf, 1995). This cooling patch can be seen in IPCC reports since 2001, in the 5$^{th}$ and 6$^{th}$ report even in the Summary for Policy Makers (Fig 1). It is one of the most striking large-scale anomalies in observed global warming, and this region coincides with where climate models have long predicted cooling as a result of a weakening AMOC (Stouffer et al., 2006; Manabe and Stouffer, 1995). While the global ocean has warmed, this region in the subpolar North Atlantic has not only resisted this trend but has cooled over more than a century, which could be explained by a decline of the AMOC's heat transport.

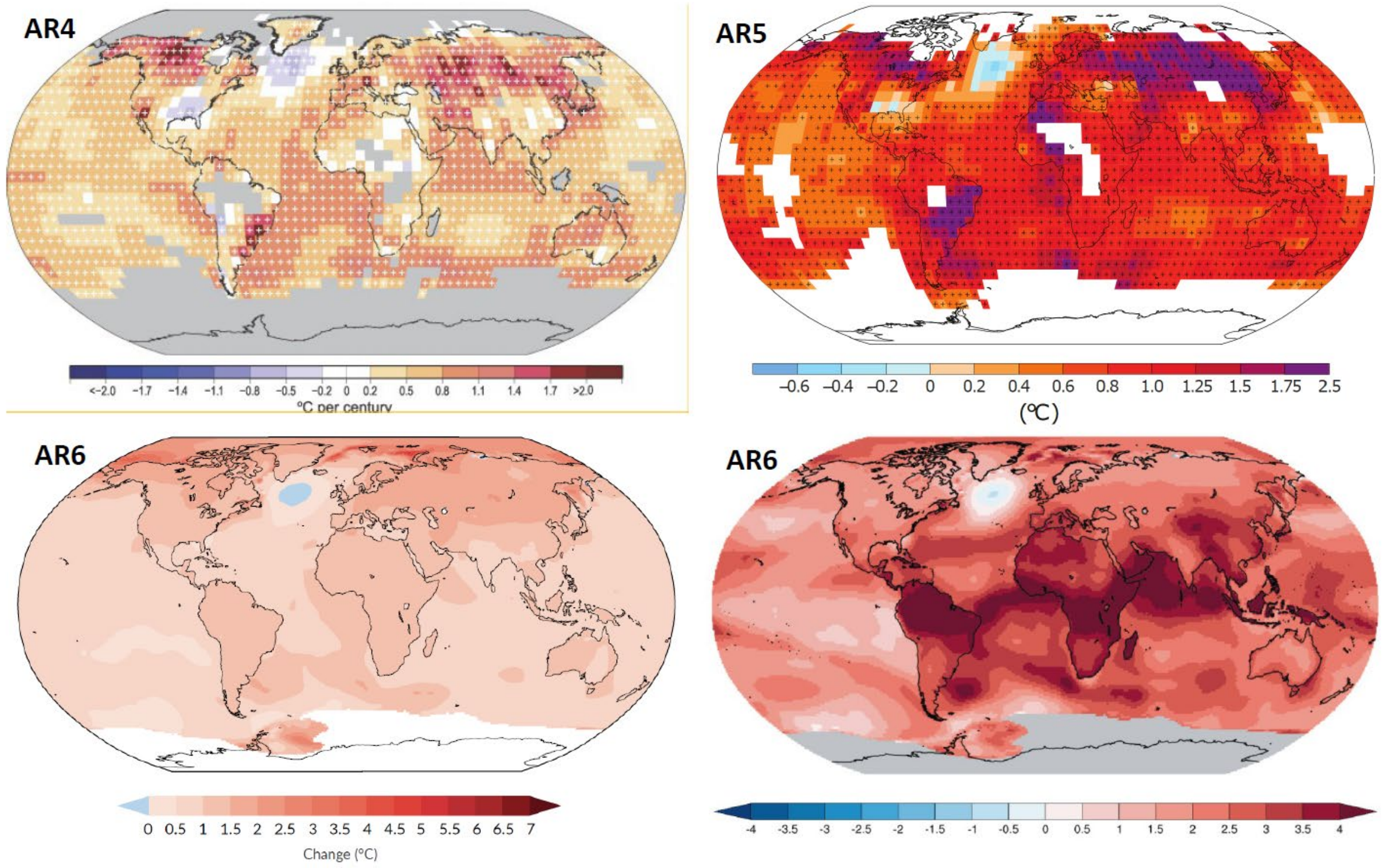

***Figure 1*** *The North Atlantic 'cold blob' in successive IPCC assessments. Near-surface air temperature change maps from AR4 (1901-2005), AR5 (1901-2012) and AR6 (1901-2020) show cooling in the subpolar North Atlantic south of Greenland. Earlier reports display the warming trend per century, while AR6 shows warming per degree of global warming (lower left) and signal-to-noise ratio (lower right). The cooling patch coincides with the region where models simulate cooling under a weakened AMOC. Adapted from IPCC AR4 WG1 Fig. 3.9, AR5 WG1 Fig. SPM.1, and AR6 WGI Fig. SPM.5.*

## 3. Temperature fingerprints

The link between this 'cold blob' and AMOC slowing is supported by both models and observations. It is most robust in the cold half-year, as in summer the 'cold blob' can be covered by a shallow mixed layer with a temperature dominated by surface forcing (England et al., 2025).

Early work demonstrated a close correspondence between subpolar Atlantic sea surface temperature and overturning strength in climate model simulations (Latif et al., 2004). Subsequent analyses have shown that the cooling in the North Atlantic is part of a larger-scale pattern, anticorrelated with warming in the South Atlantic (Dima and Lohmann, 2010) - the so-called bipolar 'seesaw' (Stocker, 1998). This pattern reflects changes in northward heat transport and is a robust feature of AMOC variability (Drijfhout et al., 2012). In a recent update to their seminal study, (Dima et al., 2021) have applied separate EOF analyses to the North and South Atlantic for the period 1854-2016, both independently pointing to a very similar time history of an AMOC in decline. (Rahmstorf et al., 2015) used the proxy surface temperature reconstruction by (Mann et al., 2008) to reconstruct AMOC strength since the year 900 AD (shown in Fig. 4). This reconstruction finds a fairly stable AMOC in the first thousand years followed by an unprecedented decline in the 20$^{th}$ Century.

In addition to the cooling in the subpolar North Atlantic, AMOC slowing leads to a northward shift of the Gulf Stream and associated warming along the North American coast, for reasons of vorticity conservation (Zhang, 2008; Zhang and Vallis, 2007). Together, these features form a characteristic spatial pattern (Fig. 2) that has been identified in both observations and model simulations and is also seen in sea surface temperature trends over 1870-2016 (Caesar et al., 2018).

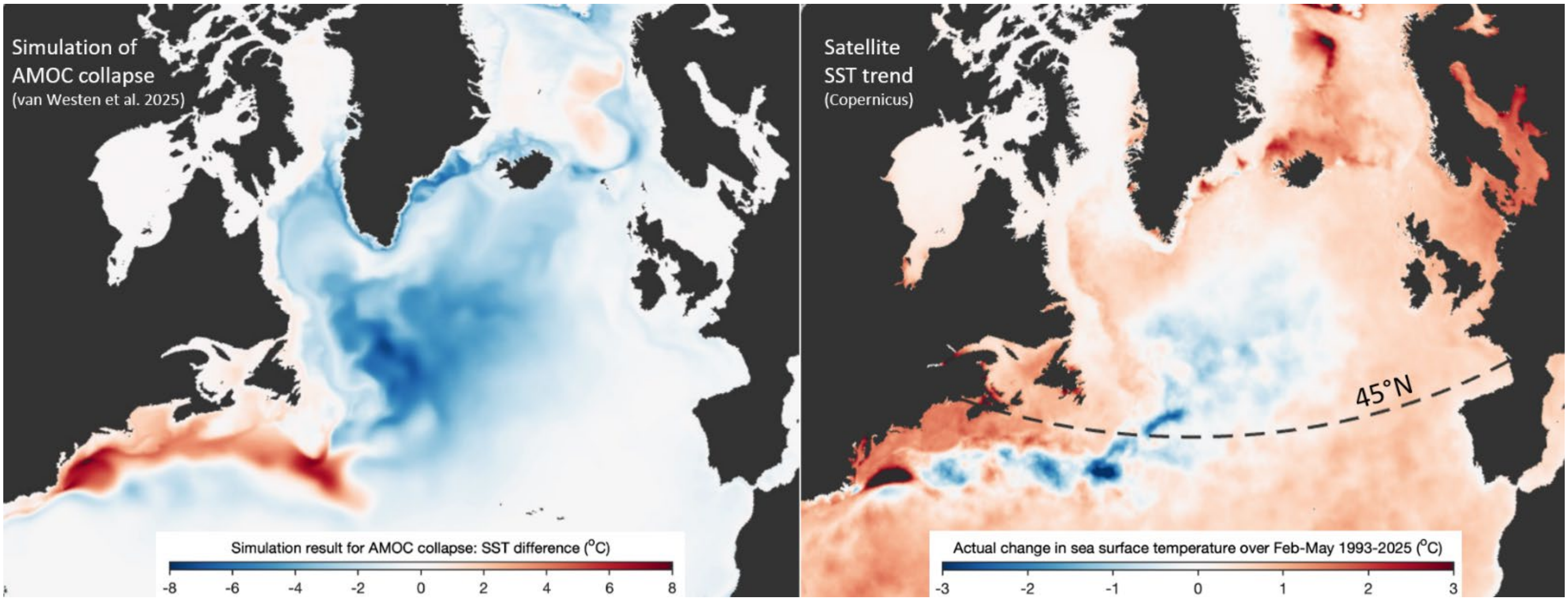

***Figure 2*** *Sea surface temperature change in a simulation of AMOC collapse in a strongly eddying ocean model (van Westen et al., 2025a) and linear trend in Copernicus satellite observations 1993-2025. The characteristic cold blob and northward Gulf Stream shift can be seen in both. The comparison is qualitative: the observations include global warming, whereas the model simulation shows purely the response to an AMOC shutdown. This 'fingerprint' pattern is also seen in SST trends since the 19$^{th}$ Century (Caesar et al., 2018). Figure by Ruijian Gou.*

Attribution studies further support the link between ‘cold blob’ and AMOC slowdown. Combining observations with climate model simulations, (Drijfhout et al., 2012) demonstrated that the warming hole is associated with AMOC changes rather than direct radiative forcing. Subsequently, (Chemke et al., 2020) showed that the warming hole has recently emerged from internal climate variability and “is indeed caused by anthropogenic emissions of greenhouse gases, and is related to changes in the oceanic circulation”. Anthropogenic aerosol pollution has warmed the ‘cold blob’ region since 1950, since its indirect effect of strengthening the AMOC more than outweighs its direct effect on surface fluxes which would cool the sea surface in the region (Dagan et al., 2020; Qasmi, 2023).

In a similar vein, (Li and Liu, 2025) “demonstrate that only models simulating a weakened historical Atlantic overturning can broadly reproduce the observed cooling and freshening in the warming hole region.” They estimate the AMOC slowing trend from 1900 to 2005 as between 0.1 and 0.3 Sv per decade, i.e. about **1-3 Sv**.

As counter-argument, (Li et al., 2021) argue with a simple 1-D energy balance at each point that a cooling trend in the subpolar Atlantic can arise by increasing surface heat loss from the ocean. However, the surface heat flux anomaly driving the SST is computed from bulk formulae which depend on the difference of atmospheric temperature and SST, where the atmospheric temperature is taken from reanalysis data. The observed cooling trend is thus included in the forcing of the passive 1-D models, so it is no surprise that they reproduce the cold blob and also the warming along the US coast.

(He et al., 2022) argue they can explain the cold blob in a slab-ocean experiment. However, the fully coupled climate-model run they start with has two cold blobs, in the Irminger Sea and in the southwest Labrador Sea, the latter caused by strong cold westerly winds from the continent and not found in observations. In the slab ocean version, the realistic cold blob disappears and the unrealistic one expands across into the Irminger Sea, but with a fraction of the amplitude observed and seen in the coupled model. (Li and Liu, 2025) have compared dynamic-ocean and slab-ocean versions of ten different climate models and found that the cold blob disappears in the slab ocean cases.

Furthermore, observation-constrained reanalysis data indicate that surface heat-flux changes dampen multidecadal heat content changes rather than causing them. Surface heat fluxes thus respond to ocean heat transport changes, with periods of strong AMOC and warm SST corresponding to large surface heat loss (Rahmstorf et al., 2026). (Li and Liu, 2025) find a >99% significant correlation between different versions of the SST fingerprint with the AMOC decline in coupled climate models.

A key advantage of SST for reconstructing anthropogenic AMOC changes is that reliable SST measurements go back to the $19^{th}$ Century, and the North Atlantic is well sampled because of the early abundance of ship traffic between Europe and North America. This long time-horizon is important for distinguishing climate trends from multidecadal variability of the AMOC.

While SST will not deliver perfect AMOC reconstructions given other influences on SST, it is unlikely that the remarkable long-term cooling trend in the subpolar Atlantic has occurred without a substantial contribution from reduced AMOC heat transport to the region.

## 4. Heat content indicators

While surface temperatures go further back in time, subsurface heat content changes have an advantage in being less sensitive to surface forcing. However, deep observations are much sparser, especially before 1950. An alternative AMOC fingerprint is based on the increasing heat content over a large region (15°S–45°N, depth range 0–2000 m) where this 'heat pile-up' has been shown to correlate with the AMOC strength in models. Based on this correlation and observational heat content data, a study diagnosed an AMOC slowdown by **1.31 ± 0.39 Sv** since the 1950s (Ren et al., 2025a). The quantitative translation into AMOC transport is based on the model-derived relationship between the indicator and AMOC strength.

A similar but more localised approach uses the fact that according to models, a weakening of the AMOC causes a warming in the equatorial Atlantic (1°S-1°N) at depths between 1,000 and 2,000 meters (Ren et al., 2025b). This study concludes that this mid-depth temperature change is a better indicator for AMOC change on decadal and longer timescales than other proxies, by being less influenced by surface forcing. They write: "Observations reveal a robust mid-depth warming since 1960 that emerged from natural variability in the early 2000s," and the study "estimates an AMOC slowdown of **~2 Sv** since the 1950s," consistent with the fingerprint studies discussed above (Figure 3).

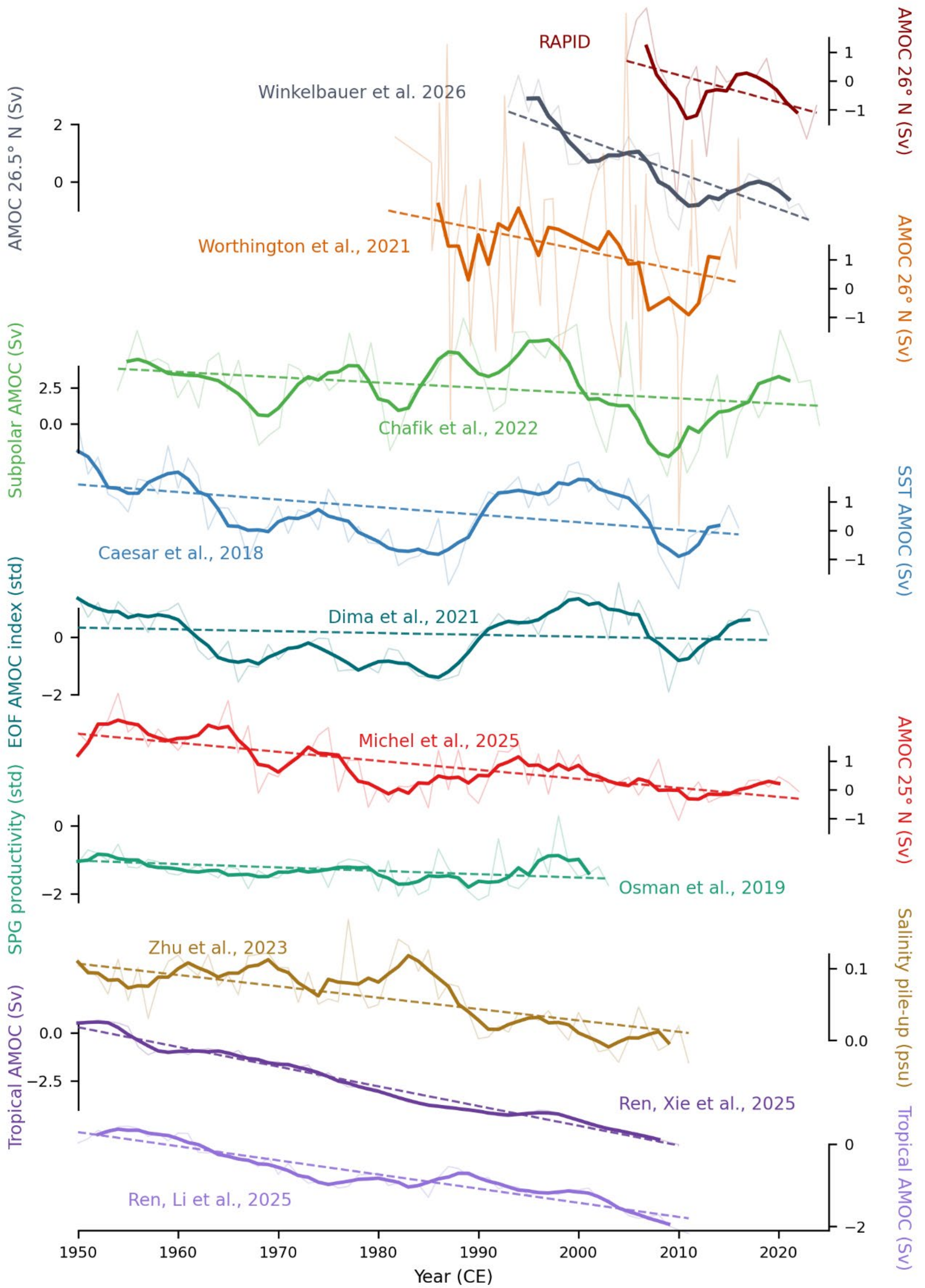
RAPID
AMOC 26° N (Sv)
Winkelbauer et al. 2026
AMOC 26.5° N (Sv)
Worthington et al., 2021
AMOC 26° N (Sv)
Subpolar AMOC (Sv)
Chafik et al., 2022
SST AMOC (Sv)
Caesar et al., 2018
EOF AMOC index (std)
Dima et al., 2021
Michel et al., 2025
AMOC 25° N (Sv)
SPG productivity (std)
Osman et al., 2019
Zhu et al., 2023
Salinity pile-up (psu)
Tropical AMOC (Sv)
Ren, Xie et al., 2025
Ren, Li et al., 2025
Tropical AMOC (Sv)
Year (CE)

***Figure 3*** *Observational estimates of AMOC variability since the mid-twentieth century. Thin lines show annual values, thick lines centred 5-year running means, and dashed lines linear trends calculated from the unsmoothed data over the full available period. The series have different units and vertical scales, so their amplitudes are not directly comparable. Some series are shifted in time to account for the expected propagation of AMOC-related signals: subpolar density (~4 years earlier), subpolar SST (~6 years later), South Atlantic salinity (~10 years later), equatorial mid-depth temperature (~10 years later), and productivity (~10 years later). Where available, shifts follow published estimates; otherwise, they are illustrative and based on physical considerations. All trends and statistical tests use the original, unshifted calendar years.*

In Table 1, we compile linear trends calculated from the unsmoothed annual values of the records shown in Figure 3. We show trends over the full available record and, where data are available, over the period since 1993, when the ocean-reanalysis ensemble and the high-resolution satellite observations shown in Fig. 2 become available. For records extending to the present, this provides an approximately three-decade comparison period. Linear trends are estimated using ordinary least-squares regression with p-values adjusted for lag-1 autocorrelation of the residuals. Note that a slowing with p-value above 0.05 indicates that the null hypothesis of no trend cannot be rejected at the conventional 95% significance threshold; that is less conclusive evidence for AMOC slowing, but it is not evidence against it (see section 12 below).

| Dataset | Unit | Full available record | | | Since 1993 | | |
|---|---|---|---|---|---|---|---|
| | | Period | Total fitted change | p (AR1-adjusted) | Period | Total fitted change | p (AR1-adjusted) |
| RAPID | Sv | Apr 2004–Mar 2024 | −1.894 | 0.170 | Apr 2004–Mar 2024 | −1.894 | 0.170 |
| Winkelbauer et al. (2026) | Sv | 1993–2023 | −3.877 | <0.001 | 1993–2023 | −3.877 | <0.001 |
| Worthington et al. (2021) | Sv | 1981–2016 | −2.560 | 0.162 | 1993–2016 | −1.216 | 0.551 |
| Michel et al. (2025) | Sv | 1950–2022 | −2.283 | <0.001 | 1993–2022 | −1.122 | 0.002 |
| Caesar et al. (2018) | Sv | 1950–2022 | −1.915 | 0.075 | 1993–2022 | −0.653 | 0.742 |
| Chafik et al. (2022) | Sv | 1950–2020 | −2.606 | 0.281 | 1993–2020 | −1.487 | 0.731 |
| Osman et al. (2019) | standardized index | 1950–2013 | −0.639 | 0.008 | — | — | — |
| Dima et al. (2021), updated series | standardized index | 1950–2025 | −0.478 | 0.678 | 1993–2025 | 0.169 | 0.879 |

| Dataset | Unit | Full available record | | | Since 1993 | | |
|---|---|---|---|---|---|---|---|
| | | Period | Total fitted change | p (AR1-adjusted) | Period | Total fitted change | p (AR1-adjusted) |
| Zhu et al. (2023) | psu | 1950–2021 | −0.114 | <0.001 | 1993–2021 | −0.106 | 0.004 |
| Ren, Xie et al. (2025) | Sv | 1957–2020 | −6.561 | <0.001 | 1993–2020 | −2.103 | — |
| Ren, Li et al. (2025) | Sv | 1960–2021 | −2.128 | <0.001 | 1993–2021 | −1.276 | 0.026 |

***Table 1.*** *Linear trends in observational AMOC estimates and indirect indicators. RAPID annual means are calculated for April–March periods. Trends are calculated consistently from unsmoothed annual values over the full available record and since 1993, or from the first available year thereafter. Trend values are reported as the total fitted change over the indicated period, calculated as the fitted trend per year multiplied by the number of years covered, in the units of the respective series. The reported p-values are adjusted for lag-1 autocorrelation of the regression residuals. All calculations use the original, unshifted calendar years. These regressions are descriptive and do not imply that the original authors interpreted each record as showing a statistically significant trend. An em dash indicates that an adjusted p-value could not be calculated.*

## 5. Salinity fingerprints

Salinity patterns provide an additional tracer of AMOC variability. (Holliday et al., 2020) found that in the cold blob region, the salinity has reached the lowest level in 120 years of data. Their study focuses on a recent freshening during 2012 to 2016 which they attribute to a wind-driven ocean circulation change, including a slowing of the North Atlantic Current and a diversion of Arctic-origin freshwater from the western boundary into the eastern basins. This mechanism can explain the short-term freshening event, but does not speak against an additional long-term contribution from reduced northward transport of warm and salty subtropical waters (Häkkinen et al., 2011). On centennial timescales, the subpolar 'fresh blob', which is in the same space as the cold blob, also comes with a salty counterpart north of the Gulf Stream, co-located with the warm part of the temperature fingerprint (Holliday et al., 2020). This spatial correspondence is consistent with a northward shift of the Gulf Stream and a slowing AMOC.

(Zhu and Liu, 2020) propose that AMOC changes can be detected in the subtropical South Atlantic in an indicator named the Atlantic 'salinity pile-up': a greater increase in basin-mean salinity in the subtropical Atlantic than in the Indo-Pacific, extending from the surface to the thermocline at approximately 300–500 m. In their analysis of 13 CMIP5 models, this pattern is not driven by local surface forcing and correlates with AMOC strength. Using this model-derived relationship, they find an observed salinity change over 1900-2017 indicative of **about 2 Sv** AMOC weakening.

In a follow-up study, (Zhu et al., 2023) argue that the South Atlantic salinity pile-up provides a remote AMOC fingerprint with a higher trend-to-variability ratio than North Atlantic fingerprints, because interdecadal AMOC variability is strongly attenuated as it propagates southward, whereas much of the centennial trend is retained. They interpret the accelerated decline of the observed salinity index since the 1990s as evidence of an accelerated AMOC weakening that began approximately one decade earlier, in the 1980s. This timing is derived from modelled relationships between the northern AMOC and the delayed South Atlantic salinity response.

The northern and southern salinity indicators therefore provide complementary evidence. Subpolar freshening is physically consistent with reduced northward transport of warm and saline waters, but is also strongly influenced by winds, internal variability and the routing of Arctic-origin freshwater. The South Atlantic salinity pile-up is less directly affected by North Atlantic interdecadal variability, but its interpretation is subject to other potential influences, including Pacific teleconnections and changes in Agulhas leakage, as well as relatively sparse early observations. Because the principal non-AMOC influences on the two indicators differ, their agreement adds weight to the overall interpretation, without constituting two fully independent lines of evidence for AMOC weakening.

## 6. Deep water renewal

The renewal and ventilation of deep water masses is a key function of the AMOC. One way to assess changes in ventilation is through the “age" of seawater, defined as the duration since its last contact with the ocean surface. This has been done for the North Atlantic for the period 1985 to 2021 by using the chlorofluorocarbon-12 and sulfur hexafluoride concentrations in seawater as abiotic tracers (Guo et al., 2026). The authors argue that water age provides a unique advantage in studying changes in ocean ventilation, including broad spatial coverage and longer time range than the RAPID measurements. Their study concludes that “the North Atlantic waters are generally aging”, which they interpret as evidence of reduced ventilation and a climate-change signal rather than natural variability.

A complementary approach for reconstructing overturning is the established WMT (Water Mass Transformation) framework. This integrates hydrographic observations of ocean temperature and salinity with atmospheric reanalyses and estimates of subsurface mixing and volume tendencies, and provides a physically grounded estimate of the overturning stream function across the Atlantic basin. A recent study applied this physics-based WMT framework to multiple observation-based ocean datasets and atmospheric reanalysis fields, finding a 13% $\pm$ 2% or **~2.4 Sv** AMOC decline at 26°N since 1980 (Li et al., 2025).

The tracer-age and WMT results thus have different methodological uncertainties and constrain different aspects of the circulation. Taken together, they provide complementary evidence consistent with reduced North Atlantic ventilation and AMOC weakening.

## 7. Density trends

Density changes are interesting since density combines the effects of temperature and salinity and can both reflect AMOC-related water-mass changes and directly affect the pressure gradients that help drive the circulation. (Chafik et al., 2022) argue that “the strong relationship between Irminger Sea density and subpolar AMOC change present in OSNAP and GloSea5 provides us with a robust, albeit indirect, measure that can be used to examine long-term AMOC trends.” They conclude that the observed Irminger Sea density decline implies an AMOC weakening of **~2.2 Sv** over the period 1950–2019, although the trend does not quite reach 95% significance in their analysis because of the large interannual and decadal variability.

The conversion of the density change into AMOC transport units is indirect and assumes that the relationship identified from the relatively short OSNAP record and supported by GloSea5 also applies on multidecadal timescales. We therefore regard the Irminger Sea density decline as physically grounded but indirect evidence of AMOC weakening. The corresponding time series is shown in Figure 3; and Table 1 reports our consistently calculated trend and AR1-adjusted p-value.

**8. Reanalysis data**

Reanalysis data sets use atmosphere and/or ocean models to analyse the past climate evolution, thereby assimilating observational data. In recent years there has been much work and substantial progress in evaluating and improving these reanalysis products. A recent study analysed the Atlantic Ocean heat transport trends for 1993–2023 in ocean reanalysis products, and found a negative trend in all 14 products included in the trend analysis at 26°N (see Fig. 3). The mean trend is negative at all latitudes from the equator to 60°N (Winkelbauer et al., 2026). This closely correlates with AMOC weakening in all the examined reanalysis products. The ensemble average heat transport reduction was found to be 0.058 ± 0.039 PW per decade and the AMOC weakening 1.23 ± 0.57 Sv per decade, corresponding to a 15% heat transport and **2-5 Sv** AMOC reduction at 26°N over the last three decades.

Atmospheric reanalysis outputs can also be used to deduce AMOC changes by computing the integral of Atlantic Ocean heat loss over a given latitude range as it changes over time (Mayer et al., 2023; Terhaar et al., 2025). Climate model simulations suggest this is dominated by AMOC changes. There are two different ways to derive the surface fluxes from the reanalysis data: (a) the fluxes computed by bulk formulae from air-sea temperature difference, wind speed etc., and (b) the fluxes derived from an atmospheric energy balance using top-of-atmosphere radiation balance and horizontal transport divergence (Trenberth et al., 2019). Method (b) is generally considered more accurate, since the energy balance calculation is constrained by observational data, the resulting global ocean energy imbalance is an order of magnitude smaller, and different reanalysis products give consistent results unlike for method (a) (Mayer et al., 2024). A study using method (a) has concluded that the AMOC shows no decline since the 1960s (Terhaar et al., 2025), while using fluxes from method (b) do reveal an AMOC weakening trend by **1.5-4.5 Sv** over the past 30 years at 26°N (Bell et al., 2026), consistent with the results of the ocean reanalysis. Overall, we conclude that the balance of evidence from reanalysis data supports an ongoing AMOC slowing.

**9. Paleoclimate proxy data**

A number of paleoclimatic studies have used different types of proxy data to reconstruct the AMOC. These were compiled and analysed statistically in (Caesar et al., 2021; Caesar et al., 2022), who concluded that after a long and relatively stable period, there was an initial AMOC weakening starting in the nineteenth century, followed by a more rapid decline starting in the mid-twentieth century. In recent decades these data indicate the weakest AMOC in at least a millennium (Figure 4). While paleoclimatic proxy data are indirect indicators which have their limitations, it is remarkable how very different reconstruction methods give rather consistent results. Some use grain size of sortable silt below the North Atlantic Deep Water flow, some use isotope data (e.g. oxygen-18) or microplankton species abundances. Those proxy series covering also the full 20$^{th}$ Century (Osman et al., 2019; Rahmstorf et al., 2015; Thornalley et al., 2018) appear consistent with observations and other reconstructions (Caesar et al., 2022). A broader compilation of proxy data reviewed by (Moffa-Sánchez et al., 2019) includes more heterogeneous records of regional North Atlantic and Arctic variability, many of which are influenced by several processes and are not intended as quantitative AMOC reconstructions. Figure 4 therefore presents a representative set of long-term AMOC reconstructions rather than a comprehensive compilation of North Atlantic proxy records.

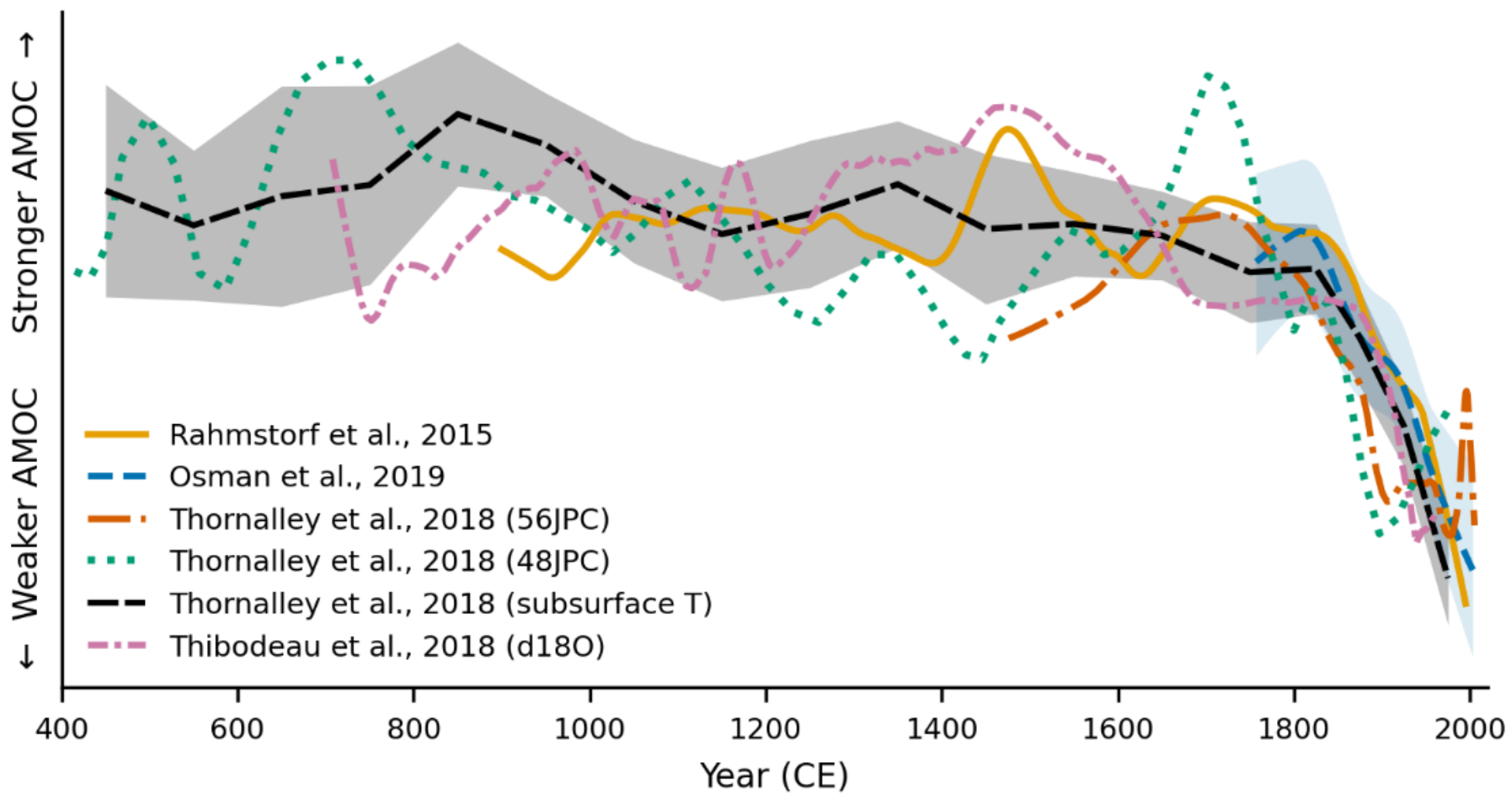

***Figure 4*** *Long-term reconstructions of AMOC variability over the past millennium. Shown are multi-proxy reconstructions (Osman et al., 2019; Rahmstorf et al., 2015), sediment-based proxies from the North Atlantic (Thornalley et al., 2018), including sortable silt (bottom current strength) and Mg/Ca-derived subsurface temperature and δ¹⁸O-based reconstruction of water masses (Thibodeau et al., 2018). Except for the subsurface-temperature stack, the records are smoothed using a 100-year LOWESS filter. All series are standardized over the displayed period and vertically offset to facilitate comparison; their amplitudes are therefore not directly comparable. For the original proxy units and further details on the individual records, see Caesar et al. (2021). The blue shading shows the 2.5th–97.5th percentile range of the Osman et al. (2019) reconstruction, and the grey shading shows the ±2 standard-error uncertainty of the Thornalley et al. (2018) subsurface-temperature stack.*

*These diverse records indicate relatively stable conditions over much of the preceding millennium, followed by a broad decline since the nineteenth to twentieth century.*

## 10. Other Approaches

As discussed above, an AMOC weakening is expected to cause a northward shift of the Gulf Stream (Figure 2) for reasons of vorticity conservation (Zhang, 2008; Zhang and Vallis, 2007). Therefore, (van Westen and Dijkstra, 2026) have used satellite altimeter data to analyse the Gulf Stream position for the period 1993–2024. They find a >95% statistically significant northward shift trend of 55 km during this time interval, not caused by wind changes and thus providing further indirect evidence for an ongoing AMOC weakening. For a longer timeseries, they analysed the position of the Gulf Stream north wall since 1965 as defined by the 15°C isotherm at 200 m depth. This metric likewise indicates a northward shift, with >99% significance.

(Michel et al., 2025) have trained a set of convolutional neural networks on climate model data in order to find and use the best SST pattern which correlates with AMOC strength at latitudes 20°-60°N at 5° intervals. They then applied this to observed SST from the HadISST dataset (Rayner et al., 2003) to reconstruct the AMOC strength for 1900-2024. At all latitudes except 35° N they find an AMOC decline, with the decline starting in mid-20$^{th}$ Century and reaching a record-weak AMOC after 2010.

These two rather different approaches provide more independent evidence that an AMOC weakening has been underway for decades.

## 11. Is there evidence against an AMOC slowing?

Several studies are often cited as evidence against an AMOC weakening. Here, it is important not to compare apples and oranges. A steady or slightly increasing inflow into and deep outflow from the Nordic Seas is sometimes cited as indication of a stable AMOC – but this is an expected response to an AMOC weakening in the subtropical North Atlantic (Roewer et al., 2026) rather than an argument against it. This complexity of the flow in higher northern latitudes is a reason why we focus on the AMOC south of 45°N in this article. Similarly, the observation-based reconstructions of (Fraser and Cunningham, 2021) at ~50°N and (Wett et al., 2023) at 47°N show no significant long-term trends. However, both are located north of 45°N, where regional overturning changes need not follow those of the AMOC farther south.Other studies find that "natural variability has dominated" (Latif et al., 2022) – but that does not mean there hasn't also been an underlying weakening. In fact, not only the variability but also the forced AMOC response identified in this study in historic model runs shows the same SST fingerprint of AMOC slowing as shown in our Fig. 2 above (see their Fig. 2).

For the South Atlantic, 'synthetic' AMOC estimates from 2002 were recently published for 25°S and 34.5°S showing no trend (Johnson et al., 2024). However, below-surface data down to 850 m depth consist of temperature from XBT casts, with salinity computed using historical T/S relationships, and below 850 m climatological temperature and salinity data were used (Dong et al., 2015). That is ill-suited for long-term trends and at 34.5°S results are in stark contrast to those of the SAMBA array (which shows an AMOC weakening, but starts only in 2009 and is thus too short to be considered here).

Some studies show an AMOC slowing but dismiss it on the grounds of it not being statistically significant (Chafik et al., 2022; Rossby et al., 2022; Worthington et al., 2021). However, evidence for a non-significant slowing is still (albeit weaker) evidence for slowing, and should not be taken as evidence against slowing, though this is "a common misinterpretation in climate science" (Shepherd, 2021). Lack of significance does not mean a trend is not real, it just means that from these data *alone*, we cannot reject a prior null hypothesis of no trend with 95% confidence.

## 12. Discussion: Combining the Evidence

None of the many indicators for AMOC discussed here provide by itself a conclusive answer to the question whether AMOC is changing in response to climate change. The gold standard data from RAPID do show a weakening trend consistent with climate model projections, but so far only for about 20 years. For the observed longer-term indicators, each one could be affected by other processes than AMOC, and model results could be wrong.

But what do the various lines of evidence tell us when considered together? It is easy to underestimate how multiple lines of evidence combine in a statistical way, and this seems to be at the core of the disagreement about AMOC slowing. To give just one illustrative example: Let us consider the AMOC trend since 1993, for which the multi-model mean clearly predicts an AMOC weakening both in CMIP5 and CMIP6 models. But let's assume we are quite sceptical about these models, so to start with we assume only a 50:50 chance that AMOC has actually slowed in the real ocean since 1993. With that prior expectation we now consider observations. For the observed cooling trend in the Irminger Sea as shown in Fig. 2, we estimate that if the AMOC has actually slowed, there is a 75% likelihood to observe such a cooling patch or 'cold blob'. But we also think that even if the AMOC has *not* slowed, there still is a 33% chance such a cold blob could arise by some other mechanism, including natural variability.

With these illustrative assumptions, basic probability maths (going back to Thomas Bayes in the 18$^{th}$ Century) then gives us an update from our 50% prior probability to a 69% posterior probability of AMOC slowing. So just by this one piece of observational evidence with uncertain link to AMOC, we find a 'likely' AMOC slowing in IPCC terms. Now if we add a second, independent line of evidence, let's say the South Atlantic salinity pile-up, and we estimate an 80% likelihood to see that if AMOC has weakened, and a 20% likelihood if not, that pushes the overall posterior probability to over 90%, making it 'very likely' in IPCC terms that AMOC has in fact weakened.

This is just an example to illustrate how even uncertain evidence adds up. A full quantitative estimate is beyond the scope of this article and more complex, given there are many more lines of evidence as discussed above, but how independent they are is hard to establish. Of course, any evidence that AMOC has increased would lower the posterior probability, but such evidence is hard to find.

Bayesian frameworks provide a coherent structure for synthesizing evidence from diverse sources and are well-established e.g. in medical science, but also increasingly in climate science. A prominent example is the climate sensitivity range given in the IPCC AR6, which is based on the Bayesian assessment of a diverse range of evidence, including feedback process understanding, the historical climate record and the paleoclimate record (Sherwood et al., 2020).

What we argue in this article is that there is a large variety of different indications of AMOC change all pointing in the same direction of slowing: direct measurements as far back as they are available, a range of ocean and atmosphere reanalysis products, multiple sea surface temperature and heat content fingerprints, different salinity fingerprints including a 'salinity pile-up' in the South Atlantic, the Irminger Sea density decline, paleoclimate data such as grain sizes below the North Atlantic Deep Water flow, tracers of reduced deep-water renewal, the northward shifting Gulf Stream, and emerging machine learning reconstructions.

It is this large number of complementary lines of evidence pointing toward an AMOC weakening which gives us high confidence in our assessment. The evidence indicates an AMOC weakening of about 1–3 Sv, with median estimates of 2.3 Sv since around 1950 and 1.4 Sv since 1993. Note, however, that the exact time spans covered by the data differ, that decadal variability is superimposed on the trends, and that the uncertainty range is substantial.

We conclude that the balance of many different lines of evidence suggests that a past and ongoing AMOC slowing may not be certain - but it is at least very likely, using the IPCC definition of that term (i.e. >90% probability). This has important implications for the future evolution of the climate system, given the central role of the AMOC in regulating heat, carbon, and freshwater transport in the Atlantic.

Risk management does not require certainty, particularly where potential consequences are large and difficult to reverse. Three priorities follow. First, since important changes are underway, the existing AMOC observing systems need to be sustained, better coordinated and expanded into a robust monitoring and early-warning system, potentially through a joint observatory supported by several countries (Judah, 2026; Nummelin et al., 2026). Second, the possibility of substantial further weakening or collapse should be considered in long-term climate adaptation and preparedness. Such AMOC changes could contribute to accelerated regional sea-level rise (Levermann et al., 2005; van Westen et al., 2025a), major changes in precipitation and drought patterns (van Westen et al., 2025b; Bellomo and Mehling, 2024), and regional cooling in north-western Europe (van Westen and Baatsen, 2025). And last but most importantly, rapid reductions in fossil-fuel use and other greenhouse-gas emissions remain essential to limit further warming and to reduce the risk of large further AMOC weakening, and of crossing a tipping point of no return. The location and timing of such a threshold remain uncertain, but this uncertainty is no reason to delay policy measures to reduce the risk.

### Code and Data Availability

The code used to process the data and generate Figures 3 and 4 is available at https://gitlab.pik-potsdam.de/caesar/Rahmstorf_and_Caesar_2026. The repository contains the full analysis workflow, including data processing, smoothing, and trend estimation routines.

The datasets used in this study originate from previously published sources and are cited in the manuscript. The majority of datasets are publicly available and are accessed via URLs provided in the code repository. For datasets that are not publicly archived, access information is provided, and these data can be obtained from the original authors upon reasonable request. The availability status of each dataset is documented in the code repository.

### Author contributions

SR conceptualised the study and led the manuscript writing. LC developed the code and performed the data analysis for Figures 3 and 4 as well as contributed to the interpretation of the results and writing of the manuscript.

### Acknowledgement

We thank Swinda Falkena for valuable suggestions.

### Competing interests

The authors declare that they have no conflict of interest.